\documentclass[twocolumn,aps,showpacs]{revtex4}

\usepackage[dvips]{graphicx,graphics}
\usepackage{amssymb,epsfig}

\begin{document}
\title{Gradual transition from insulator to semimetal of Ca$_{1-x}$Eu$_{x}$B$_{6}$ with increasing Eu concentration}
\author{R.R. Urbano$^1$, P.G. Pagliuso$^1$, C. Rettori$^1$, P. Schlottmann$^{2}$, J.L. Sarrao$^3$, A. Bianchi$^3$, S. Nakatsuji$^{2,4}$, Z. Fisk$^{2,5}$, E. Velazquez$^6$ and S.B. Oseroff$^6$}

\affiliation{$^1$ Instituto de F\'{\i}sica "Gleb Wataghin", UNICAMP, Campinas-SP, 13083-970, Brazil.\\
$^2$ Department of Physics and National High Magnetic Field Laboratory, Florida State University, Tallahassee, FL 32306, U.S.A.\\
$^3$ Los Alamos National Laboratory, Los Alamos, New Mexico 87545, U.S.A.\\
$^4$ Department of Physics, University of Kyoto, Kyoto 606-8502, Japan\\
$^5$ Department of Physics, University of California, Davis, CA 95616, U.S.A.\\
$^6$ San Diego State University, San Diego, California 92182,
U.S.A.}

\date{\today}

\begin{abstract}

The local environment of Eu$^{2+}$ ($4f^{7}$, $S=7/2$) in
Ca$_{1-x}$Eu$_{x}$B$_{6}$ ($0.003\leqslant x\leqslant 1.00$) is
investigated by means of electron spin resonance (ESR). For
$x\lesssim 0.003$ the spectra show resolved \textit{fine} and
\textit{hyperfine} structures due to the cubic crystal
\textit{electric} field and nuclear \textit{hyperfine} field,
respectively. The resonances have Lorentzian line shape,
indicating an \textit{insulating} environment for the Eu$^{2+}$
ions. For $0.003\lesssim x\lesssim 0.07$, as $x$ increases, the
ESR lines broaden due to local distortions caused by the Eu/Ca
ions substitution. For $0.07\lesssim x\lesssim 0.30$, the lines
broaden further and the spectra gradually change from Lorentzian
to Dysonian resonances, suggesting a coexistence of both
\textit{insulating} and \textit{metallic} environments for
the Eu$^{2+}$ ions. In contrast to Ca$_{1-x}$Gd$_{x}$B$_{6}$, the
\textit{fine} structure is still observable up to $x\approx 0.15$.
For $x\gtrsim 0.30 $ the \textit{fine} and \textit{hyperfine}
structures are no longer observed, the line width increases, and
the line shape is purely Dysonian anticipating the
\textit{semimetallic} character of EuB$_{6}$. This broadening is
attributed to a spin-flip scattering relaxation process due to the
exchange interaction between conduction and Eu$^{2+}$ $4f$
electrons. High field ESR measurements for $x\gtrsim 0.15$ reveal
smaller and anisotropic line widths, which are attributed to
magnetic polarons and Fermi surface effects, respectively.
\end{abstract}

\maketitle

\section{\textbf{INTRODUCTION}}

The system Ca$_{1-x}$R$_{x}$B$_{6}$ (R = rare-earths, specially
La) has become the focus of extensive scientific investigations
since weak-ferromagnetism (WF) at high-temperature ($T_{c}
\approx $ $600-800$ K) has been reported in these materials
by Young {\it et al.}\cite{Young} Over the last few years,
enormous efforts were devoted, both theoretically
\cite{Zhitomirsky,Leon,Victor,Jarlborg,Ceperley,Tromp,Massidda}
and experimentally,
\cite{Young,Urbano,Moriwaka,Hall,Vonlanthen,Ott,Gavilano,Denlinger,Giannio,Rhyee}
to establish the origin of the WF in Ca$_{1-x}$La$_{x}$B$_{6}$
and its relationship with the actual conducting nature of R doped
CaB$_{6}$. However, the nature of the WF of the parent compound
CaB$_{6}$ is still controversial. Studies of the de Haas-van
Alphen effect,\cite{Moriwaka,Hall} the plasma edge in optical
spectroscopy,\cite{Vonlanthen,Ott} and some electrical resistivity
measurements \cite{Souma} support a semimetallic character for
CaB$_{6}$, whereas NMR,\cite{Gavilano} thermopower,\cite{Giannio}
angle-resolved photoemission (ARPES),\cite{Denlinger} and a
different set of resistivity measurements \cite{Vonlanthen,Rhyee}
suggest that CaB$_{6}$ is a well defined semiconductor.
High-resolution ARPES by Souma {\it et al.} \cite{Souma} revealed an
energy gap of about 1 eV between the valence and conduction bands
and a carrier density of the order of $5\times 10^{19}$ cm$^{-3}$
for their CaB$_{6}$ single crystals. Vonlanthen {\it et al.}
\cite{Vonlanthen} reported that, depending on the crystal growth
method, undoped CaB$_{6}$ can also show WF. They argue that
self-doping attributed to defects might occur. Terashima
{\it et al.}\cite{Terashima} reported that data for
Ca$_{0.995}$La$_{0.005}$B$_{6}$ is strongly sample dependent and,
lately, doubts about the intrinsic nature of the WF in these
systems are being raised.\cite{Matsubayashi} It has also been
argued that CaB$_{6}$ is a $\sim 1$ eV-gap semiconductor and
that the intrinsic WF could be hidden by the ferromagnetism (FM)
of Fe and Ni impurities at the surface of the crystals.\cite{Bennett}

The electron spin resonance (ESR) study on Gd$^{3+}$ in
Ca$_{1-x}$Gd$_{x}$B$_{6}$ by Urbano {\it et al.}\cite{Urbano} has
shown that the doping process in this compound leads to an
inhomogeneous material. Coexistence of \textit{metallic} and
\textit{insulating} local environments for Gd$^{3+}$ were
observed for Gd concentrations of $\sim 1000$ ppm.\cite{Urbano}
No evidence of WF was found in the Gd$^{3+}$ ESR spectra of either
\textit{metallic} or \textit{insulating} regions. Nonetheless,
some of the crystals displayed a $g\sim 2.00$ narrow ($\Delta
H\approx 10$ Oe) resonance, which disappeared after a gentle
etching of the sample. These results
suggest that the Gd$^{3+} $ $4f$-electrons are shielded from the
WF field, i.e. the WF might be confined to small regions or to the
surface of the sample away from the impurity R sites. In addition,
for Gd concentrations $\gtrsim $ $2000$ ppm, an intriguing
collapse of the cubic crystal field (CF) \textit{fine} structure
was observed in the ESR spectra of Gd$^{3+}$ ions in the
\textit{metallic} regions.\cite{Urbano}

In contrast to CaB$_{6}$, EuB$_{6}$ is a well established
\textit{semimetallic} material that orders ferromagnetically at
$T_{c}\approx 15$ K,\cite{Sullow,Rhyee,Paschen} although recently
Wigger {\it et al.} \cite{G.A. Wigger} interpreted their EuB$_{6}$ data
in the framework of a small-gap semiconductor. The electronic
configuration of Eu$^{2+}$ ions ($4f^{7}$, $S=7/2$) is identical
to that of the Gd$^{3+}$ ions. However, the effect of Gd$^{3+}$
and Eu$^{2+}$ doping in CaB$_{6}$ is quite different since
Eu$^{2+}$ has the same valence as Ca$^{2+}$, while Gd$^{3+}$
delivers an extra electron to the system creating a hydrogen-like
donor state with large Bohr radius. The \textit{insulator} to
\textit{metal} transition revealed by the change in the ESR line
shape is then reached when the Gd donor bound-states overlap and
start to form a \textit{percolative} network. Since not all
Gd-sites participate in this network, a coexistence of
\textit{metallic} and \textit{insulating} local environments are
observed for Gd concentrations of $\sim $ 1000 ppm.

The substitution of Ca$^{2+}$ by Eu$^{2+}$ impurities does not
yield a donor bound state. Instead, the broken translational
invariance of the lattice introduces a localized split-off state
from the valence/conduction band. The energy of such state lies in
the gap of the semiconductor and its spatial extension is of the
order of one unit cell. Thus an impurity band for Eu$^{2+}$ only
forms at much higher concentrations than for Gd$^{3+}$, as it is
indeed observed in our experiments.

Therefore, an ESR study, probing the local Eu$^{2+}$ environment
in Ca$_{1-x}$Eu$_{x}$B$_{6}$, is of great interest to understand
the magnetic/non-magnetic and metallic/non-metallic properties of
these materials. In this work we present a systematic Eu$^{2+}$
ESR study of Ca$_{1-x} $Eu$_{x}$B$_{6}$ single crystals for
$0.003\leqslant x\leqslant 1.00$. Preliminary X band data in some
of these samples were already presented previously.\cite{Jmmm} For
EuB$_{6}$ ($x$ = 1.00) Urbano \textit{et al.}\cite{R. Urbano}
have recently attributed the broad line width observed in their
ESR experiments to a spin-flip scattering relaxation process due
to the exchange interaction between the Eu$^{2+}$ $4f$ and
conduction electrons. As a consequence, the observed field,
temperature, and angular dependence of the ESR line width could be
associated with the Fermi surface of the conduction states and the
formation of magnetic polarons.

\section{\textbf{EXPERIMENTS}}

Single crystals of Ca$_{1-x}$Eu$_{x}$B$_{6}$ ($0.003\leqslant
x\leqslant 1.00$) were grown as described in Ref. \cite{Young}.
The structure and phase purity were checked by x-ray powder
diffraction and the crystal orientation determined by Laue x-ray
diffraction. Most of the ESR experiments were done on $\sim
1\times 0.5\times 0.3$ mm$^{3}$ single crystals in a Bruker
spectrometer using a X-band ($9.479$ GHz) TE$_{102}$ room-$T$
cavity and a Q band ($34.481$ GHz) cool split-ring cavity, both
coupled to a $T$-controller using a helium gas flux system for
$4.2 K \lesssim T\lesssim 300$ K. $M(T,H)$ measurements for $2$
K $\lesssim $ $T$ $\lesssim 300$ K were taken in a Quantum Design
SQUID-RSO $dc$-magnetometer. The Eu$^{2+}$ concentration was
obtained from Curie-Weiss fits of the susceptibility data.

\section{\textbf{EXPERIMENTAL RESULTS}}

Figure 1 presents the X-band ESR spectrum of Eu$^{2+}$ in a
Ca$_{1-x}$Eu$_{x}$B$_{6}$ single crystal for $x=0.003$ at room-$T$
for $H\parallel [001]$. The spectrum shows the \textit{fine} and
\textit{hyperfine} structures corresponding to seven groups
($4f^{7},S=7/2$) of twelve hyperfine resonances due to the
$^{151}$Eu$^{2+}$ (47.8\%; $I=5/2$) and $^{153}$Eu$^{2+}$ (52.2\%;
$I=5/2$) isotopes. The line shape of the individual resonances is
Lorentzian as expected for an \textit{insulating} environment for
the Eu$^{2+}$ ions, in agreement with the spectrum reported for
$x\lesssim 0.001$ (see Fig.~2b in Ref. \cite{Urbano}). The angular
dependence of the \textit{fine} structure (seven groups of twelve
hyperfine resonances) for the field rotated in the ($110$) plane
is shown in the inset of Fig.~1. This angular dependence was
simulated using the spin Hamiltonian $H=g\beta \mathbf{H}\cdot
\mathbf{S}+(b_{4}/60)(\mathbf{O}_{4}^{0}+5\mathbf{O}_{4}^{4})+
(b_{6}/1260)(\mathbf{O}_{6}^{0}-21\mathbf{O}_{6}^{4})$. The first
term is the Zeeman interaction and the second and third ones are
the cubic CF terms.\cite{Bleaney} The isotropic hyperfine
coupling, $A\mathbf{I}\cdot \mathbf{S}$, is not included in the
simulation, which yielded the following values for the spin
Hamiltonian parameters: $g=1.988(4)$, $b_{4}=-38.5(5)$ Oe, and
$b_{6}\lesssim 2(1)$ Oe. The hyperfine parameters,
$^{151}A=35.2(4)$ Oe and $^{153}A=15.8(4)$ Oe, were obtained from
simulations of twelve hyperfine resonances within the $1/2
\leftrightarrow -1/2$ spin transition of the spectrum for
$H\parallel [001]$.

Figure 2 presents the X-band ESR spectra of Eu$^{2+}$ in
Ca$_{1-x}$Eu$_{x}$B$_{6}$ single crystals with $0.003\leqslant
x\leqslant 1.00$ at room-$T$ and $H\parallel [001]$. The data show
that the individual resonances and the spectra, as a whole, become
broader as $x$ increases. Nonetheless, the Eu$^{2+}$ resolved
\textit{fine} structure is still observed up to Eu concentrations
of the order of $x\approx 0.15$. This is in contrast to Gd$^{3+}$
in Ca$_{1-x}$Gd$_{x}$B$_{6}$, where a Gd concentration of just
$\sim 0.1-0.2$\% is sufficient to collapse the entire spectrum
into a single \textit{metallic} narrow line.\cite{Urbano} The
\textit{fine} structure for the crystals with $x=0.07$, $0.10$ and
$0.15$ also show the angular dependence corresponding to a CF of
cubic symmetry. The corresponding spin Hamiltonian parameters are
given in Table I.

Figures 3a and 3b display the room-$T$ angular dependence of the X
band ESR peak-to-peak line width, $\Delta H$, for the sample with
$x=0.15$ for the field rotated in the $(110)$ and $(100)$ planes,
respectively. $\Delta H$ was estimated by fitting the spectra to a
single Dysonian line that \textit{averages} the shape of the
spectra. Although this method yields only a rough estimate of $\Delta
H$, it provides an accurate account of the relative changes in
$\Delta H$.
The angular variation observed with X band
corresponds to a CF of cubic symmetry.\cite{Davidov,Rettori,Luft}
The solid lines are the fitting of the data to $\Delta
H=a+b\bigm|1-5\sin ^{2}(\theta )+\frac{15}{4}\sin ^{4}(\theta
)\bigm|$ and $\Delta H=a+b\bigm|1-\frac{5}{4}\sin ^{2}(2\theta
)\bigm|$ for the $(110)$ and $(100)$ planes,
respectively.\cite{Bleaney} The fitting parameters are $a=276(5)$
Oe and $b=72(5)$ Oe. For $x\gtrsim 0.30$ the spectra show a single
Dysonian resonance with no resolved \textit{fine} structure. For
these crystals the angular dependence of $\Delta H$ presents a
minimum at $[111]$ when the field is rotated in the $(110)$ plane,
and a minimum along the $[110]$ direction when the field is
rotated in the $(100)$ plane. This behavior is shown in Figs. 3c
and 3d for $x=0.30$ and indicates that the origin of the line
width anisotropy is not due to a CF of cubic symmetry. Similar
behavior was found for the samples with $x=0.60$ (not shown here)
and $x=1.00$ (see Ref.\cite{R. Urbano} and Figs. 6 and 7 below).
The solid lines are the fitting of the data to $\Delta
H^{2}(\theta ,\phi )=A+Bf_{4}(\theta ,\phi )+Cf_{6}(\theta ,\phi
)$ for the angular dependence in the $(110)$ and $(100)$ planes.
The functions $f_{4}(\theta ,\phi )$ and $f_{6}(\theta ,\phi )$
are the linear combinations of spherical harmonics of fourth and
sixth order having cubic symmetry.\cite{OsoCalvo} The parameters
$A$, $B$ and $C$ depend on $H$ and $T$, and for the present data
we found $A=423(2)$ kOe, $B=2.3(5)$ kOe and $C=-0.03(2)$ kOe.

Figures 4a, 4b and 4c display the Q-band spectra at room-$T$ for
the $x=0.07,0.10,$ and $0.15$ samples of Fig. 2. Since the fits in
Figs. 3a and 3b using a single line are not completely satisfactory,
we simulate the experimental spectra in Fig. 4 as the superposition
of two different Eu$^{2+}$ ESR spectra: a spectrum with resolved
\textit{fine} structure of Lorentzian resonances (\textit{f}SL)
corresponding to Eu$^{2+}$ ions in \textit{insulating} media, and
a Dysonian ($D$) resonance associated to Eu$^{2+}$ ions in a
\textit{metallic} environment. For $x\gtrsim 0.30$, however, the
spectra are well fitted by a single Dysonian resonance with
nearly the same $g$-value and increasing line width as $x$
increases (see Fig. 4d). The fits in Figs. 4a, 4b and 4c are a
crude simulation for the coexistence of two different local
environments for the Eu$^{2+}$ ions in the region $0.07\lesssim
x \lesssim 0.15.$

Figures 5a and 5b present the angular dependence of $\Delta H$ in
the $(110)$ plane for the $x=0.10$ and $0.15$ samples at room-$T$
measured with X and Q bands. The data is analyzed as in Figs. 3a and
3b in terms of a single resonance. The angular dependence indicates
that, for these two samples, there are two competing contributions
to $\Delta H$, one due to the unresolved CF \textit{fine} structure
of cubic symmetry and a second one of lower symmetry which is more
evident for Q band. The solid lines are fittings of $\Delta H$
using a weighted superposition of these two contributions, i.e.
$\Delta H= \tilde{a}+\tilde{b}\bigm|1-5\sin ^{2}(\theta )+
\frac{15}{4}\sin^{4}(\theta )\bigm|+[\tilde{A}+\tilde{B}f_{4}(\theta
,\phi)+\tilde{C}f_{6}(\theta ,\phi )]^{1/2}$. The fitting parameters
are given in Table II. Within the accuracy of the measurements,
the samples with $x\leqslant 0.07$ did not show significant
differences in the \textit{fine} structure spectra between X and
Q bands.

The angular dependence of $\Delta H$ measured at X and Q bands in
the $(110)$, $(100)$ and $(001)$ planes is shown for $x=0.30$ at
297 K and 50 K in Fig. 6, and for $x=1.00$ at 297 K and 150 K in
Fig. 7. Note that (1) $\Delta H$ becomes smaller and more anisotropic
at high fields (Q band), and (2) within the accuracy of the
measurements, the narrowing and anisotropy is nearly independent
of temperature for $T \gtrsim 100$ K.

For the same samples of Fig. 2, Fig. 8 displays the room-$T$ X
band ESR spectra for the magnetic field in the (110) plane along
the angle of minimum $\Delta H$, i.e. $\sim 30^{\circ}$ ($\sim
55^{\circ}$) away from the [001] direction for $x\leqslant 0.15$
($x\geqslant 0.30$). The data show the increase of $\Delta H$ as
$x$ increases. For $x$ between 0.07 and 0.15 the line shape can be
approximated as an admixture of Lorentzian and Dysonian (see Fig. 4).
For $x\geqslant 0.30$ the line shape is purely Dysonian with a
$A/B\approx 2.3$ ratio corresponding to a skin depth much smaller
than the size of the crystals.\cite{Pake} The Dysonian line shape
suggests a \textit{metallic} environment for the Eu$^{2+}$ ions,
confirming an increasing \textit{semimetallic} character of
Ca$_{1-x}$Eu$_{x}$B$_{6}$ as $x$ increases.

Figure 9a displays $\Delta H$ for the spectra of Fig. 8 measured
at X and Q bands. For the Eu$^{2+}$ ions in the \textit{metallic}
environment, $\Delta H$ becomes narrower at higher fields (Q
band). Fig. 9b shows the line width anisotropy, $^{Q,X}\Delta
H_{[001]}-$ $^{Q,X}\Delta H_{\min}$, which is larger for the Q
band than for the X band. Fig. 9c shows that the line width
reduction, $^{Q}\Delta H_{\min}-^{X}\Delta H_{\min}$, is nearly
$x$-independent in the metallic regime (see Figs. 5, 6 and 7).
This behavior suggests that the Eu$^{2+}$-Eu$^{2+}$ dipolar
interaction cannot be responsible for the concentration broadening
of the Eu$^{2+}$ resonance shown in Figs. 2 and 8.\cite{Sperlich}

Figures 10a and 10b present, respectively, the $T$-dependence of
$\Delta H$ and the $g$-value measured at X band. Here $H$ lies in
the (110) plane and is oriented along the angle of minimum $\Delta H$,
i.e. $\sim 30^{\circ}$ ($\sim 55^{\circ}$) away from the [001]
direction for $x\leqslant 0.15$ ($x=0.30$ and $0.60$). Note that
above $\sim 50$ K and for $x\geqslant 0.30$ both, $\Delta H$ and
$g$ are $T$-independent. There is a dramatic increase in both $g$
and $\Delta H$ for $x$ between $0.15$ and $0.30$. The broadening
of the resonance and the increase of the $g$-value for $x\gtrsim
0.30$ and $T\lesssim 30$ K indicates that there are short range
ferromagnetic correlations already in the paramagnetic phase
($T\gtrsim T_{c}$).

The $g$-values and CF parameters, $b_{4}$, measured for the low
concentration crystals, are listed in Table I. As usual for
insulators, we found a $T$-dependent $b_{4}$. For $x=0.003$ we
measured a decrease of about $5$\% in $|b_{4}|$ when the
temperature is raised from $\sim 10$ K to $\sim 300$ K.
This $T$-dependence is actually expected from the thermal
expansion of the lattice.\cite{Buckmaster,Bleaney,Bartkowski}
In addition, within the accuracy of the measurements $|b_{4}|$
for $x=0.15$ is about the same as for $x=0.003$. This is also
not unexpected, since the lattice parameter does not significantly
change with $x$.

\section{\textbf{ANALYSIS AND DISCUSSION}}

The spectra of Eu$^{2+}$ in Ca$_{1-x}$Eu$_{x}$B$_{6}$ presented in
this paper can be classified into three different concentration
regimes. (1) For $x<0.07$ the resonance line shapes are Lorentzian
and, therefore, the local environment of the Eu$^{2+}$ sites is
\textit{insulating}. As a consequence of the relatively slow
spin-lattice relaxation, the \textit{fine} and \textit{hyperfine}
structures are fully resolved in this regime. From the isotropic
$g$-value ($\approx 1.988(4)$) and the anisotropy of the \textit{fine}
structure it is inferred that the local symmetry is cubic (see
Fig. 1). (2) For $0.07\leqslant x\leqslant 0.15$ the ESR spectra
can be interpreted as a superposition of a resolved \textit{fine}
structure (\textit{f}SL) and a Dysonian ($D$) resonance (see Fig.
4). The angular dependence of the \textit{averaged} line width
corresponds to a superposition of two competing contributions,
one due to an unresolved CF of cubic symmetry and a second one
with overall cubic symmetry due to three equivalent tetragonal
contributions along the three [001] axes (see Figs. 2, 3, 4 and 5).
The $g$-values are isotropic, within the accuracy of our measurements.
In this regime the ESR line shape begins to show a Dysonian shape,
i.e. it starts to display \textit{metallic} character (see Fig. 8).
Here the microwave skin depth became comparable to the size of our
crystals ($\sim 1\times 0.5\times 0.3$ mm$^{3}$). (3) For $x
\geqslant 0.30$ the Eu$^{2+}$ ESR line shape is purely Dysonian,
i.e., there is spin diffusion and the local environment
is \textit{metallic}. The \textit{fine} and \textit{hyperfine}
structures can no longer be resolved and the $g$-value and $\Delta
H$ are $T$-independent down to $\sim 30$ K (see Figs. 10a and
10b).

The absence, or negligible, linear $T$-broadening of $\Delta H$
(\textit{Korringa} relaxation) for $x\gtrsim 0.3$ is a consequence
of the very low carrier density in these systems. Also, a strong
$q$-dependence of the exchange interaction, $J(q)$, ($q =
|k_{f}-k_{i}|$, conduction electron momentum transfer) leading to
a small average exchange interaction over the Fermi surface,
$\langle J^{2}(q) \rangle _{FS} $, could in part be responsible
for this behavior. The larger $g$-values observed for the samples
with $x\geqslant 0.30$ also suggest a strong $q$-dependence of the
exchange interaction.\cite{D.Davidov} At low $T$ a strong shift
and broadening can be observed due to ferromagnetic short range
Eu$^{2+}$-Eu$^{2+}$ correlations (see Figs. 10a and 10b). The
minimum in $\Delta H$ at $\sim 55^{\circ }$ when the field is
rotated in the $(110)$ plane and at $45^{\circ}$ when it is
rotated in the $(100)$ plane (see Figs. 3c and 3d) suggests the
presence of simultaneous tetragonal components along each of the
three $[001]$ axes, that average to a cubic symmetry. However,
within the accuracy of our measurements, a weak trigonal component
can not be excluded. This lower symmetry of the Eu$^{2+}$ site is
observed in the line width.

In the small $x$ limit, each Eu$^{2+}$ represents a charge neutral
substitution, which gives rise to a bound state in the gap of the
semiconductor, as a consequence of the broken translational
invariance. The impurity states are localized within the extension
of about an unit cell. As the number of impurity states increases
with $x$, they start to overlap and eventually form a
\textit{percolative} network. The critical concentration for
nearest neighbor site \textit{percolation} on a simple cubic
lattice is $x_{c}=0.307$.\cite{Essam} The \textit{percolation}
threshold is reduced to $x_{c}=0.137$ if next-to-nearest neighbors
are included, which correspond to neighbors in the
$[110]$-directions. Third neighbors are along the diagonals of the
cube, but this direction is blocked for the wave functions because
of the large B$_{6}^{2-}$ anions. From our results, it is
reasonable to assume that the transition from \textit{insulator}
to \textit{metal} occurs at an Eu concentration of about $14$\%.
Hence, assuming a homogeneous distribution of Eu, for $x<x_{c}$
the system is \textit{insulating} and the \textit{fine} and
\textit{hyperfine} structures could be resolved. The data at
room-$T$ shows a gradual broadening of the individual Eu$^{2+}$
resonances with increasing $x$ (see Figs. 2 and 4). We argue that
this cannot be attributed to Eu$^{2+}$-Eu$^{2+}$ magnetic
correlations because the $g$-value and $\Delta H$ are
$T$-independent (see Figs. 10a and 10b for $x<x_{c}$). Thus, the
broadening for $x \leqslant 0.15$ may be attributed to site
symmetry breaking due to Ca/Eu substitution. This inhomogeneous
broadening is probably responsible for hiding the Eu$^{2+}$
\textit{fine} and \textit{hyperfine} structures as $x$ increases.
Weak or no $T$-dependence is expected from this inhomogeneous
broadening.

For $x>x_{c}$, on the other hand, the system is \textit{metallic}
and spin-diffusion can take place, giving rise to a Dysonian line
shape. With increasing $x$ the impurity band gradually smears the
semiconducting gap at the $X$-points of the Brillouin zone and the
system evolves to a \textit{semimetal} for $x=1$
(EuB$_{6}$).\cite{Souma} EuB$_{6}$ orders FM at $T_{c}\approx
15.3$ K,\cite{Sullow,Oseroff} indicating the existence of short
and long range Eu$^{2+}$-Eu$^{2+}$ magnetic correlations around
$T_{c}$. At $T_{c}$, EuB$_{6}$ undergoes a transition into a
\textit{metallic} state at low $T$ with an increase of the number
of carriers and/or a decrease of their effective
mass.\cite{Rhyee,Paschen,Aronson} This transition is believed to
be associated with the formation of magnetic polarons, i.e. the
spin of a conduction electron polarizes the Eu$^{2+}$-spin in its
neighborhood and drags this polarization cloud as it moves. Note
that due to the \textit{semimetallic} character there are many
more Eu$^{2+}$-spins than conduction electrons. The magnetic
correlations increase as $T$ is reduced, so that the size of the
polarization clouds increases, they eventually overlap and a FM
state is obtained. The transition into a FM state is foreseen in
the ESR spectra (see Fig. 10). The $g$-values and $\Delta H$ are
independent of $T$ down to $50$ K for all $x$. However, for
$x\gtrsim 0.30$ and $T\lesssim 50$ K there is a significant
$T$-dependence. The $g$-value strongly increases due to the FM
correlations (polarons and FM short-range order). These
correlations clearly increase with $x$ and decreasing $T$. $\Delta
H$ also increases when the short-range order sets in due to the
generation of magnons.

For crystals with $x\gtrsim 0.30$ the angular dependence of
$\Delta H$ has an overall cubic symmetry that corresponds to a
superposition of three tetragonal environments. There are two
possible scenarios to explain these findings. Within the first
scenario, the origin for the tetragonal symmetry may be associated
with the Eu/Ca substitution. This scenario assumes that the CF of
an Eu$^{2+}$ ion is determined by the nearest neighbor cation
ions. If an Eu ion is surrounded by five Eu and one Ca, the local
symmetry is tetragonal. Since the Ca ion can be along any one of
the axis, there is a superposition of tetragonal symmetries along
the three directions. The overall symmetry is then cubic and the
spectra inhomogeneously broadened by three tetragonal angular
dependence. Also, more than two Ca neighbors can give rise to a
trigonal component in the line width, which was not observed. If
such component is present its value is much smaller than the
tetragonal contribution. However there is a concern with this
scenario. This mechanism does not account for the angular
dependence of stoichiometric EuB$_{6}$. It would require a much
larger number of B$_6$ vacancies than the claimed for these
samples to explain the tetragonal dependence of $\Delta
H$.\cite{FiskOseroff}

In a second possible scenario, we consider the relaxation of the
Eu$^{2+}$-spins into the conduction electron bath and the
concomitant spin-diffusion. The conduction electrons occupy three
small ellipsoidal pockets centered at the $X$-points of the
Brillouin zone, i.e., at $(\pm \pi /a,0,0)$, $(0,\pm \pi /a,0)$
and $(0,0,\pm \pi /a)$,\cite{Goodrich} where $a$ is the lattice
constant. The drift of the diffusion is then predominantly into
the direction of the major axis of the ellipsoids, i.e. along one
of the axis of the cube. Thus, each relaxation process gives rise
to a tetragonal angular dependence of $\Delta H$. The
superposition of the relaxation into the three directions is then
cubic. This mechanism explains why there is a tetragonal
dependence in the \textit{metallic} regime but not in the
\textit{insulating} one. Finally, the mechanism also applies to
stoichiometric EuB$_{6}$. For EuB$_{6}$, Urbano {\it et al.} \cite{R.
Urbano} have attributed the broad ESR line to a homogeneous
resonance, where the main contribution to $\Delta H$ involves a
spin-flip scattering relaxation process due to the exchange
interaction between the conduction and Eu$^{2+}$ $4f$-electrons.

The exchange $J$ of the order of $0.15$ eV is much larger than the
thermal energy and locally binds the conduction electrons to the
Eu$^{2+}$ spins. The relaxation process is then essentially
$T$-independent for $T > 50$ K. The exchange $J(r)$ strongly
decreases with the distance between the conduction electron and
Eu$^{2+}$ ion. At lower $T$, the thermal energy is less than the
exchange with more distant Eu ions, thus allowing the formation of
larger polarons, which eventually give rise to ferromagnetism. The
$H$, $T$ and angular-dependence of the measured $\Delta H$ lead to
the conclusion that magnetic polarons and Fermi surface effects
dominate in the spin-flip scattering of EuB$_{6}$.

Experimental support for the second scenario comes from the ESR
data and the magnetoresistance. Figs. 5, 6 and 7 show that the
high field ESR spectra present narrower and more anisotropic lines
for $x\gtrsim 0.30$, indicating that $\Delta H$ is predominantly
determined by a spin-flip scattering process. This situation is
similar to EuB$_{6}$. Furthermore, the negative magnetoresistance
observed in Ca$_{1-x}$Eu$_{x}$B$_{6}$ for $x\gtrsim 0.30$ also
supports the presence of magnetic polarons in these
systems.\cite{Paschen,G.A. Wigger,G.Wigger} Moreover, even in the
percolative region, $0.10\lesssim x\lesssim 0.15$, our data
support this second scenario. The results of Fig. 5 clearly show
that there are two competing contributions to the Eu$^{2+}$ ESR
line width. One corresponds to the {\it inhomogeneous} broadening
of unresolved CF \textit{fine} structure of cubic symmetry and the
other one is associated with the {\it homogeneous} broadening
caused by the spin-flip scattering. The latter has components of
tetragonal symmetry along the three axis with an overall cubic
dependence (see above). As $x$ increases, the homogeneous
spin-flip scattering contribution to $\Delta H$ starts to overcome
the inhomogeneous broadening due to CF effects. For $x\gtrsim
0.30$, the mobility of the carriers increases with increasing $x$
(a more connected network of impurity bound states), enhances the
spin-diffusion and contributes to the homogeneous concentration
broadening of $\Delta H$. In a recent electron microscopy study
Wigger {\it et al.} \cite{G.Wigger} found that their $x\approx
0.27$ sample presented separated regions rich in Ca$^{2+}$
(\textit{insulating}) and Eu$^{2+}$ ions (\textit{semimetallic}).
According to these results, we may associate the two contributions
to $\Delta H$ with regions rich in Ca$^{2+}$ and Eu$^{2+}$ ions,
respectively. In other words, we associate the \textit{f}SL in
Fig. 4 with regions rich in Ca$^{2+}$ and the $D$ resonance with
regions rich in Eu$^{2+}$, although in reality a distribution of
resonances, rather than just two, should be considered.

The situation is actually quite similar to the case of
Ca$_{1-x}$Gd$_{x}$B$_{6}$, where coexistence of \textit{insulating}
and \textit{metallic} regions were inferred from the Gd$^{3+}$ ESR
spectra, although, at much lower concentrations.\cite{Urbano}
Therefore, the collapse of the CF \textit{fine} structure found in
Ca$_{1-x}$Gd$_{x}$B$_{6}$ is also observed in the
Ca$_{1-x}$Eu$_{x}$B$_{6}$ system.

With respect to the controversial scenario of the conductivity of
stoichiometric CaB$_{6}$, the Eu$^{2+}$ ESR results in
Ca$_{1-x}$Eu$_{x}$B$_{6}$ for $x<0.07$ and those in
Ca$_{1-x}$R$_{x}$B$_{6}$ (R = Gd$^{3+}$, Er$^{3+}$) for $x\lesssim
0.001$,\cite{Urbano} reveal an \textit{insulating} local
environment for the dopants, Eu$^{2+}$, Gd$^{3+}$ and Er$^{3+}$,
supporting that pure CaB$_{6}$ is a ``wide-gap'' semiconductor.
Regarding the reported sample dependence for $R$-doped CaB$_{6}$,
we would like to point out that in our study neither the Eu$^{2+}$
ESR spectra nor the $M(H)$ data in Ca$_{1-x}$Eu$_{x}$B$_{6}$ were
found to be sample dependent. However, there is a strong sample
dependence in the ESR spectra of (Gd, Er)-doped CaB$_{6}$ for
concentrations of $\sim 0.1-0.2$\%. This can probably be attributed
to the donor states with large extension provided by the Ca/Gd and
Ca/Er substitution.

All our crystals presented a WF component $\lesssim 0.5$ emu/mole,
i.e., smaller than those for La-doped CaB$_{6}$ crystals grown
using the same method.\cite{Young,Urbano} Furthermore, for
$x\gtrsim 0.10$, the WF component became difficult to measure
due to the large $M$ of Eu$^{2+}$. The above results suggest that
the WF is probably caused by self-doping or extrinsic impurities
which are inherent to the employed materials and crystal growth
techniques.

\section{\textbf{CONCLUSIONS}}

In summary, we report ESR results in Ca$_{1-x}$Eu$_{x}$B$_{6}$.
As a function of Eu concentration, an evolution from
\textit{insulating} to \textit{semimetallic} character is observed
from the change in the line shape of the Eu$^{2+}$ ESR spectra.
The gradual transition between these two regimes is estimated to
occur at $0.10\lesssim x\lesssim 0.20$, indicating that a
next-to-nearest neighbor \textit{site percolation} network of
Eu$^{2+}$ bound states is the origin of this evolution. In analogy
to Ca$_{1-x}$Gd$_{x}$B$_{6}$ \cite{Urbano} a coexistence of
\textit{insulating} and \textit{metallic} regions was found for Eu
concentrations in the \textit{percolative} transition regime.
However, the collapse of the CF \textit{fine} structure is found
at a much larger of Eu concentration, because, in contrast to
Gd$^{3+}$, Eu$^{2+}$ does not provide donor carriers. The dramatic
variation of $\Delta H$ and the $g$-value at the percolative transition
revealed by our ESR data is attributed to change in the electron
mobility at these concentrations. From the ESR data for $x\leqslant
0.15$ we extracted the Eu$^{2+}$ spin Hamiltonian parameters.

For crystals with Eu concentration above the \textit{percolative}
region ($x\gtrsim 0.15$) the line width shows: (\textit{i}) a field
narrowing, which arises from magnetic polarons,\cite{R. Urbano}
(\textit{ii}) an angular dependence that corresponds to the
superposition of three tetragonal components along the $[001]$ axes,
and is related to the relaxation mechanism and the Fermi
surface,\cite{R. Urbano} and (\textit{iii}) a broadening with the
Eu concentration, which is attributed to an enhanced spin-flip
relaxation caused by the increased mobility of the carriers.
(\textit{iv}) For $x\gtrsim 0.30$ and $T\lesssim 50$ K, the
broadening and shift of the ESR spectra anticipate the onset
of FM correlations between the Eu$^{2+}$ ions due to magnetic
polarons. All measured Ca$_{1-x}$Eu$_{x}$B$_{6}$ crystals presented
a WF smaller than $\sim 0.5$ emu/mole, i.e., much smaller than
that reported by Young {\it et al.} \cite{Young} for La doped CaB$_{6}$.

The work at UNICAMP is supported by FAPESP and CNPq, and the work at the
NHMFL by NSF Cooperative Agreement No. DMR-9527035 and the State of Florida.
The support by NSF (grants Nos. DMR-0102235 and DMR-0105431) and DOE (grant
No. DE-FG02-98ER45797) is also acknowledged.


\begin{table}[!h]
\caption{X-band room-$T$ spin Hamiltonian parameters. The $g$
values were measured at the minimum line width. The $b_{4}$
parameters were obtained for $H\parallel [001]$.}
\begin{center}
\begin{tabular}{ccccc}
\hline
$x$( \% ) &    & $g$-value &  & $b_{4}$ (Oe)\\
\hline
0.3    &    & 1.988(4)  &  & -38.5(5) \\
7.0    &    & 1.997(4)  &  & -38.6(5) \\
10.0   &    & 2.001(5)  &  & -39.3(5) \\
15.0   &    & 2.003(5)  &  & -40.9(5) \\
30.0   &    & 2.03(2)   &  & - \\
60.0   &    & 2.03(2)   &  & - \\
100.0  &    & 2.03(3)   &  & - \\
\hline
\end{tabular}
\end{center}
\label{Table I}
\end{table}


\begin{table}[!h]
\caption{Room-$T$ parameters obtained from the fittings of the
angular dependence of the ESR line width in Fig. 5.}
\begin{center}
\begin{tabular}{cccccccccccccc}
\hline
$x$ ( \% )& &Band&  &   & $\tilde{a}$ (Oe) &  & $\tilde{b}$ (Oe) &  & $\tilde{A}$ (Oe)&  & $\tilde{B}$ (Oe) &  & $\tilde{C}$ (Oe) \\
\hline
  10.0 & & X  &  &   & 245(10)  &  & 110(5)  &  & 100(20)  &  &    8(3)  &  &  3(1)    \\
       & & Q  &  &   & 170(10)  &  &  60(6)  &  & 1400(100)  &  &   260(20)  &  &  1(1)    \\
  15.0 & & X  &  &   & 275(10)  &  &  50(5)  &  & 190(20)  &  &   50(5)  &  &  3(1)    \\
       & & Q  &  &   & 170(10)  &  &  30(3)  &  &  2740(30)  &  &    560(10)  &  &  8(2)    \\ \hline
\end{tabular}
\end{center}
\label{Table II}
\end{table}


\begin{figure}[!h]
\begin{center}
\includegraphics[width=1\columnwidth]{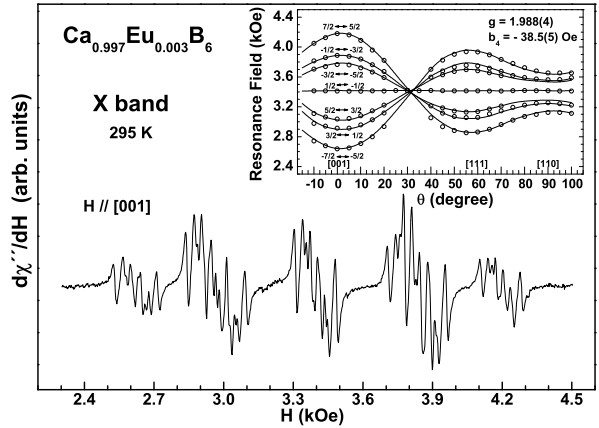}
\end{center}
\caption{X band ESR spectrum of Eu$^{2+}$ in a
Ca$_{0.997}$Eu$_{0.003}$B$_{6}$ single crystal at room-$T$ with
$H\parallel [001]$. Inset: \textit{Fine} structure angular
dependence (open circles) and simulation (solid lines) in the
$(110)$ plane.} \label{Fig.1}
\end{figure}


\begin{figure}[!h]
\begin{center}
\includegraphics[width=1\columnwidth]{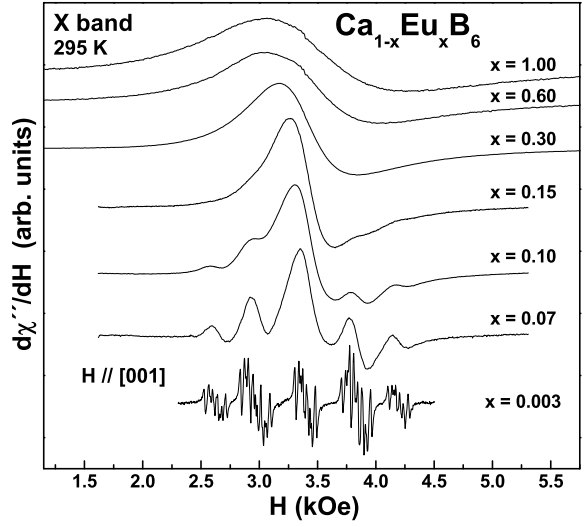}
\end{center}
\caption{ESR spectra of Eu$^{2+}$ in Ca$_{1-x}$Eu$_{x}$B$_{6}$
single crystals for $0.003\leqslant x\leqslant 1.00$ at room-$T$
for $H\parallel [001]$.} \label{Fig.2}
\end{figure}


\begin{figure}[!h]
\begin{center}
\includegraphics[width=0.8\columnwidth]{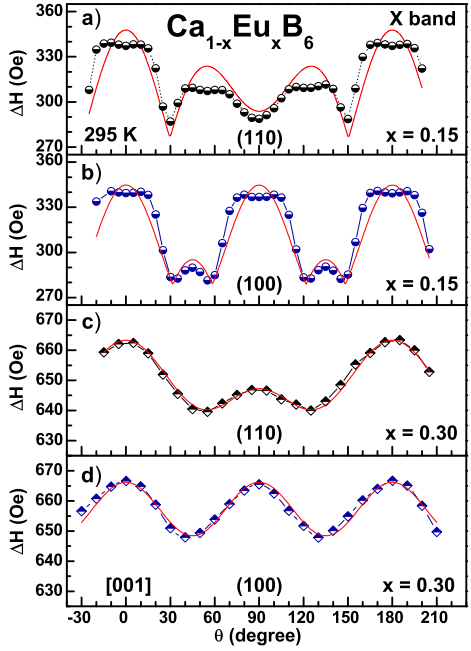}
\end{center}
\caption{(color online) Angular dependence of $\Delta H$ at
room-$T$ for $x=0.15$ and $0.30$. An "averaged" single Dysonian
line shape was used to determine $\Delta H$. For a) and c) the
angular dependence is in the $(110)$ plane, for b) and d) in the
$(001)$ plane. The solid lines are fittings to $\Delta H=a+b\
|1-5\ \sin ^{2}(\protect\theta )+\frac{15}{4}\sin
^{4}(\protect\theta)|$ for a), $\Delta H=a+b\ |1-\frac{5}{4}\ \sin
^{2}(2\ \protect\theta )|$ for b), and $\Delta
H^{2}(\protect\theta ,\protect\phi )=A+Bf_{4}(\protect\theta
,\protect\phi )+Cf_{6}(\protect\theta ,\protect\phi )$ for c) and
d) (see text). The dashed lines are a guide to the eye.}
\label{Fig.3}
\end{figure}


\begin{figure}[!h]
\begin{center}
\includegraphics[width=0.8\columnwidth]{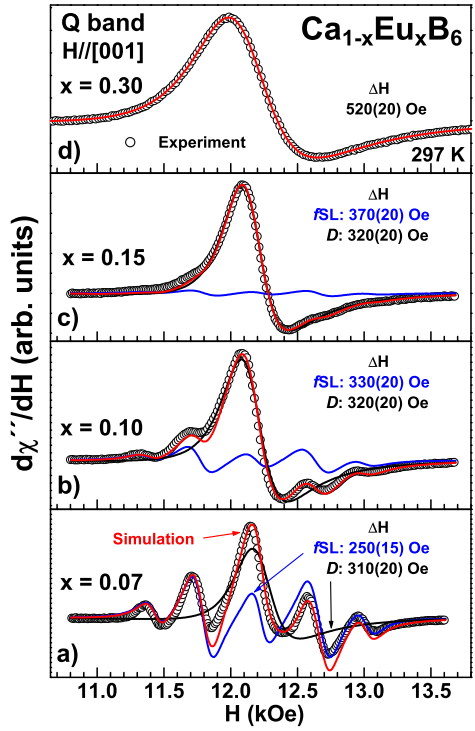}
\end{center}
\caption{(color online) Q-band ESR spectra for a) $x=0.07$;  b)
$x=0.10$; c) $x=0.15$; and d) $x=0.30$ at room-$T$ and $H\parallel
[001]$. The open symbols are the experimental data; \textit{f}SL
corresponds to the cubic \textit{fine} structure spectrum of 7
resonances ($\Delta H$ is the line width of each resonance) and
$D$ to the Dysonian resonance. In a), b) and c), the simulations
of the data are given by the superposition of \textit{f}SL and
$D$. In d) the simulation is given by a single Dysonian resonance
(see text). The $g$ values were found to be $g = 2.00(3)$ in all
cases.} \label{Fig.4}
\end{figure}


\begin{figure}[!h]
\begin{center}
\includegraphics[width=0.8\columnwidth]{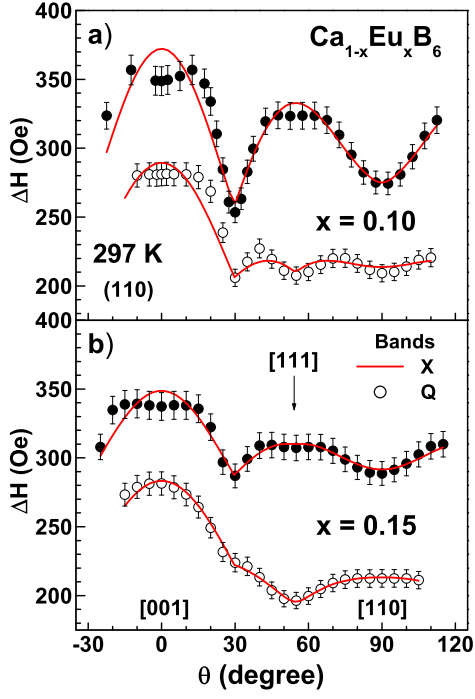}
\end{center}
\caption{(color online) X- and Q-band angular dependence of
$\Delta H$ at room-$T$ in the (110) plane: a) for $x=0.10$, and b)
$x=0.15$. An "averaged" single Dysonian line shape was used to
determine $\Delta H$. The solid lines are fittings using $\Delta
H=\tilde{a}+\tilde{b} \bigm|1- 5\sin
^{2}(\theta)+\frac{15}{4}\sin^{4}(\theta)\bigm|+
[\tilde{A}+\tilde{B}f_{4}(\theta,\phi)+\tilde{C}f_{6}(\theta
,\phi)]^{1/2}$ (see text).} \label{Fig.5}
\end{figure}


\begin{figure}[!h]
\begin{center}
\includegraphics[width=0.8\columnwidth]{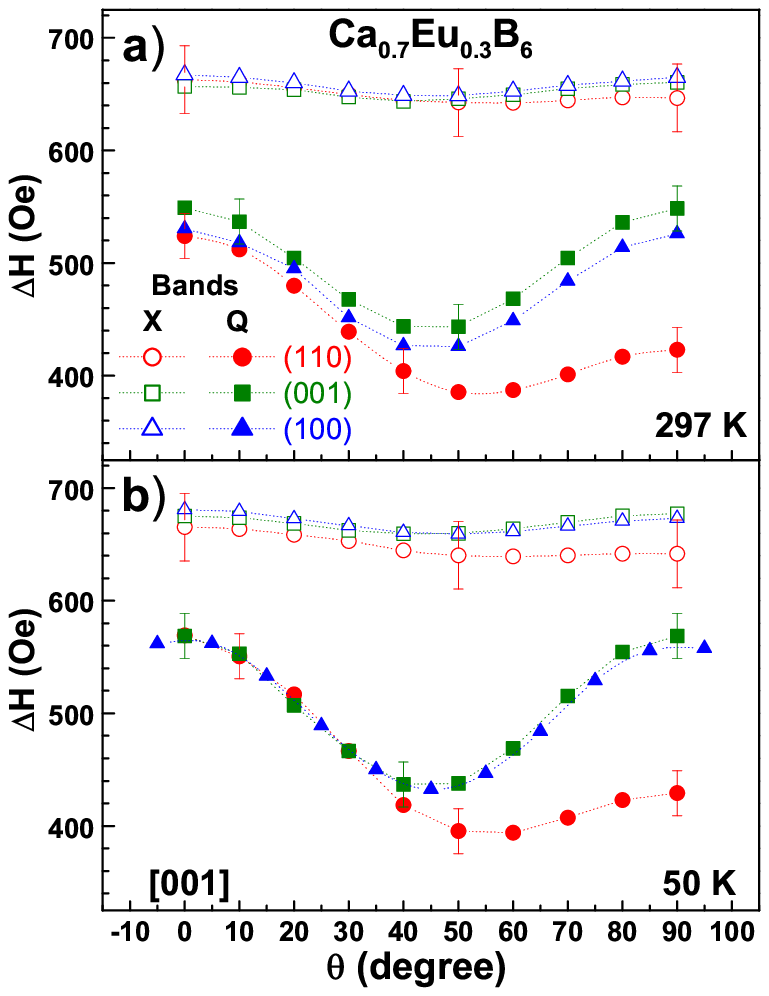}
\end{center}
\caption{(color online) Angular dependence of $\Delta H$ for X-band
(open symbols) and Q-band (full symbols) in the $(110),(001)$ and
$(100)$ planes for Ca$_{0.7}$Eu$_{0.3}$B$_{6}$ at a) room-$T$
and b) $50$ K. } \label{Fig.6}
\end{figure}


\begin{figure}[!h]
\begin{center}
\includegraphics[width=0.8\columnwidth]{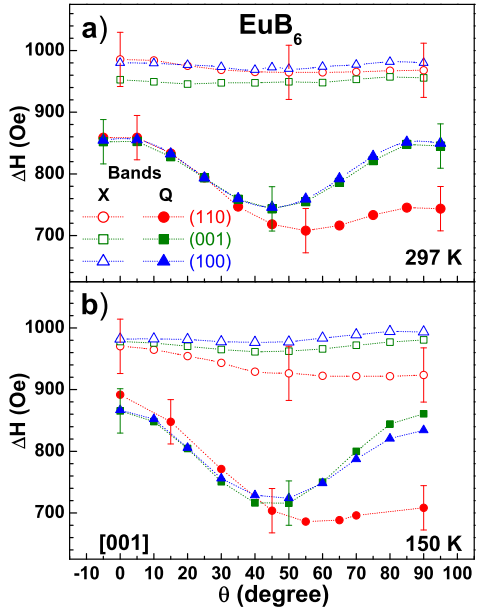}
\end{center}
\caption{(color online) Angular dependence of $\Delta H$ at X-band
(open symbols) and Q-band (full symbols) in the $(110),(001)$ and
$(100)$ planes for EuB$_{6}$ at a) room-$T$ and b) $150$ K. } \label{Fig.7}
\end{figure}


\begin{figure}[!h]
\begin{center}
\includegraphics[width=1\columnwidth]{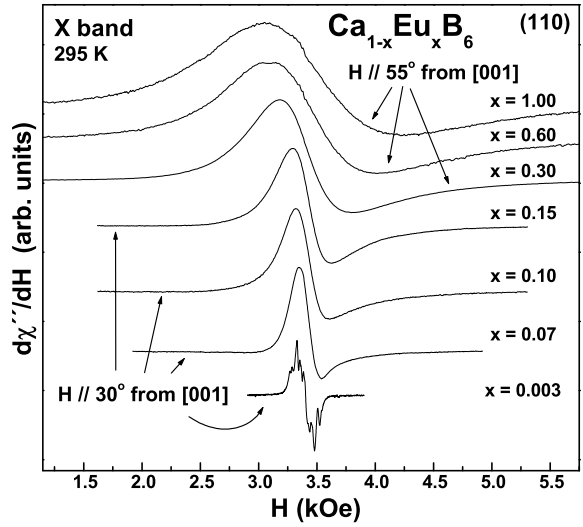}
\end{center}
\caption{ESR spectra of Eu$^{2+}$ in Ca$_{1-x}$Eu$_{x}$B$_{6}$
single crystals at room-$T$ with $H$ in the ($110$) plane. For
$0.003\leqslant x\leqslant 0.15$, $H$ is at $30^{\circ }$ from
$[001]$, while for $0.30\leqslant x\leqslant 1.00$, $H$ is at
$55^{\circ }$ from $[001]$.} \label{Fig.8}
\end{figure}


\begin{figure}[!h]
\begin{center}
\includegraphics[width=0.9\columnwidth]{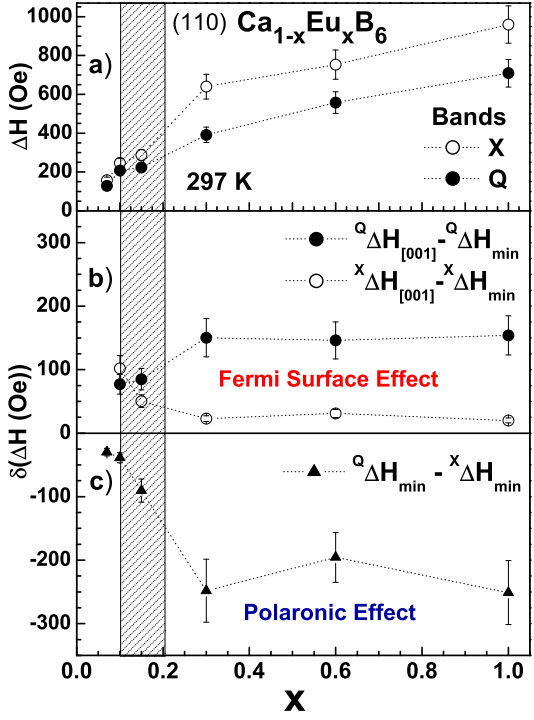}
\end{center}
\caption{(color online) $H$ and $x$-dependence of $\Delta H$ at
room-$T$: a) $^{X,Q}\Delta H$ for the spectra of Figure 8, b)
difference between the maximum and minimum line widths,
$^{Q,X}\Delta H_{[001]}-^{Q,X}\Delta H_{\min}$, and c) difference
between the minimum line widths, $^{Q}\Delta H_{\min}-$
$^{X}\Delta H_{\min}$. The gradual transition region between
\textit{insulating} and \textit{semimetallic} environments is
shown as a shaded area ($0.10\lesssim x\lesssim 0.20$).}
\label{Fig.9}
\end{figure}


\begin{figure}[!h]
\begin{center}
\includegraphics[width=1\columnwidth]{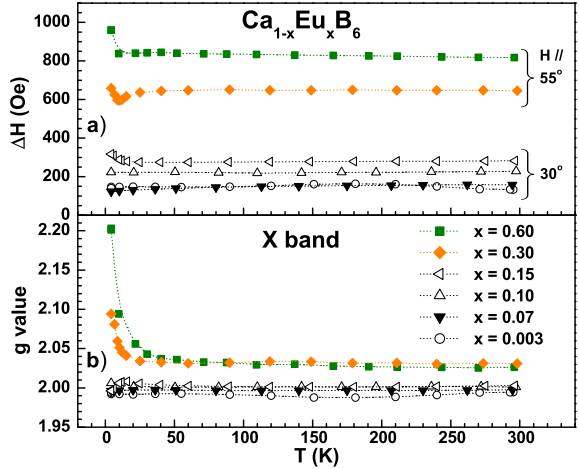}
\end{center}
\caption{(color online) X-band $T$-dependence of: a) the line-width
$\Delta H$, and b) the $g$-value, for the crystals and orientations
corresponding to Figure 8.} \label{Fig.10}
\end{figure}


\end{document}